\documentclass[twocolumn,tighten]{aastex63}
\usepackage{graphicx}
\usepackage{color}
\usepackage{amsmath}
\usepackage{mathrsfs}
\bibpunct{(}{)}{;}{a}{}{,}

\shorttitle{The Effect of Weak Cosmic Ray Heating Events on the Desorption of $\rm H_2$}
\shortauthors{Sipil\"a et al.}

\graphicspath{{./}{figures/}}

\begin{document}

\title{The Effect of Weak Cosmic Ray Heating Events on the Desorption of $\rm H_2$}

\correspondingauthor{Olli Sipil\"a}
\email{osipila@mpe.mpg.de}

\author[0000-0002-9148-1625]{Olli Sipil\"a}
\affiliation{Max-Planck-Institut f\"ur Extraterrestrische Physik, Giessenbachstrasse 1, 85748 Garching, Germany}

\author[0000-0003-1572-0505]{Kedron Silsbee}
\affiliation{University of Texas at El Paso, El Paso, TX, 79968, United States of America}
\affiliation{Max-Planck-Institut f\"ur Extraterrestrische Physik, Giessenbachstrasse 1, 85748 Garching, Germany}

\author{Naomi Carbajal}
\affiliation{University of Texas at El Paso, El Paso, TX, 79968, United States of America}

\author[0000-0003-1481-7911]{Paola Caselli}
\affiliation{Max-Planck-Institut f\"ur Extraterrestrische Physik, Giessenbachstrasse 1, 85748 Garching, Germany}

\author[0000-0003-2303-0096]{Marco Padovani}
\affiliation{INAF -- Osservatorio Astrofisico di Arcetri, Largo E. Fermi 5, 50125 Firenze, Italy}

\begin{abstract}
The typical amount of molecular hydrogen (${\rm H_2}$) in interstellar ices is not known, but significant freeze-out of ${\rm H_2}$ on dust grains is not expected. However, chemical models ubiquitously predict large amounts of $\rm H_2$ freeze-out in dense cloud conditions, and specialized treatments are needed to control the $\rm H_2$ population on grains. Here we present a numerical desorption model where the effect of weak heating events induced by cosmic rays (CRs) that heat grains to temperatures of a few tens of Kelvin at high frequencies is included, improving upon earlier desorption models that only consider strong heating events (maximum grain temperature close to 100\,K) that occur at a low frequency. A temperature of a few tens of Kelvin is high enough to induce efficient desorption of $\rm H_2$, but we find that even the weak heating events do not occur often enough to lead to significant $\rm H_2$ desorption. Taking the weak heating events into account does affect the predicted abundances of other lightly-bound species, but the effect is restricted to low column densities. We make here the canonical assumption that the grains are spherical with a radius of 0.1\,$\mu$m. It is conceivable that in the case of a grain size distribution, weak heating events could provide a boost to $\rm H_2$ desorption coming off small grains, which are the most numerous. Further studies are still required to better quantify the role of CRs in the desorption of $\rm H_2$ and other weakly bound species.
\end{abstract}


\section{Introduction}

Although molecular hydrogen (${\rm H_2}$) is by far the most abundant molecule in space, it has not been securely detected in interstellar ices to date. One detection has been claimed by \citet{Sandford93} in the infrared spectrum of WL5 in the $\rho$~Oph molecular cloud, implying a column density of $2.5\times10^{18}\,\rm cm^{-2}$ for ${\rm H_2}$ frozen in water-rich ice, which is a factor of a few larger than that measured for CO. However, this tentative detection remains unconfirmed, and the lack of secure detections of $\rm H_2$ absorption features in interstellar ices since then implies that the relative amount of $\rm H_2$ in the ice is in reality rather low. 

Conventional chemical models that describe the gas-grain chemical interaction however predict a very large amount of $\rm H_2$ accreting onto grains especially in dense clouds, and indeed a very high abundance of $\rm H_2$ in the ice is an issue in such models since decades \citep[see, e.g.,][]{Hasegawa93b}. $\rm H_2$ has a low binding energy of approximately 500\,K on water ice \citep[e.g.,][]{Katz99}, but at ambient temperatures of $\sim$10\,K, even thermal desorption is not efficient enough to remove significant quantities of it from the dust grains. The nondetection of strong $\rm H_2$ ice features in observations thus implies that an efficient means of removing $\rm H_2$ from grain surfaces, not taken into account in conventional chemical models, must be in effect.

Ways to remove excess $\rm H_2$ from grain surfaces in simulations have been discussed in the literature. For example, it has been suggested that the $\rm H_2$ population in the ice can be self-regulated because the nonpolarity of $\rm H_2$ allows species deposited on top of $\rm H_2$ to desorb much easier than they would on water ice \citep[e.g.,][]{Garrod11, Taquet14}. In these models, binding energies are varied time-dependently via a scaling relation as a function of the relative amount of $\rm H_2$ in the ice. Such scaling methods are however quite imprecise because binding energies on surfaces other than $\rm H_2O$ and CO are generally not known (some specific adsorbate-surface pairs have been studied by, e.g., \citealt{Vidali91}; \citealt{Hornekaer05}; \citealt{Cuppen09}; \citealt{Molpeceres21}). Another possible means of keeping the icy $\rm H_2$ population in check is the encounter desorption mechanism proposed by \citet{Hincelin15}, where an $\rm H_2$ molecule diffusing on the grain surface can be desorbed easily if it wanders into a binding site already occupied by $\rm H_2$. Despite these efforts, the actual mechanism(s) controlling the amount of $\rm H_2$ in the ices in space remain unknown. We note that good constraints on the $\rm H_2$ ice abundance are sorely needed also from the chemical perspective; laboratory experiments have shown that $\rm H_2$ plays a role in the formation of many molecules in the ice on grain surfaces \citep[see, e.g.,][]{Chuang18b,Martin-Domenech20,Martin-Domenech24}.

In this paper, we explore the role of cosmic ray induced desorption (hereafter CRD) as a potential new solution to the issue of excess grain-surface $\rm H_2$. We present a new numerical treatment of CRD applied to a gas-grain chemical model where, for the first time, high-frequency low deposited energy cosmic ray (CR) impacts (hereafter weak heating events) are considered simultaneously with low-frequency high deposited energy CRs (strong heating events). Existing CRD models only treat the strong heating events, and as such may be missing a significant contributor to desorption; weak heating events can raise the temperature of interstellar grains often to a few tens of Kelvin, which should a priori provide a significant boost to the desorption of lightly bound molecules, such as $\rm H_2$. We run gas-grain chemical simulations using the new CRD treatment for a set of models representing physical conditions across a molecular cloud, from the outer, low-density regions all the way to dense cores within. Although the main focus of our study is to quantify the effect of weak heating events on the $\rm H_2$ abundance in interstellar ices, we also investigate the effect on other common ice constituents such as CO, and discuss the potential implications on gas-phase chemistry.

The paper is organized as follows. Section~\ref{s:model} describes the numerical implementation of the new CRD treatment. We present the results of our simulations in Sect.\,\ref{s:results} and discuss their implications in Sect.\,\ref{s:discussion}. Our conclusions are laid out in Sect.\,\ref{s:conclusions}. Appendices~\ref{a:heatingFrequencyDerivation}~and~\ref{a:P18H} include additional discussion supporting the main text.

\section{Model}\label{s:model}

\subsection{Numerical description of CRD}

The efficiency of CRD is controlled by a competition between the cooling time of the grain ($\tau_{\rm cool}$) following a heating event and the time interval between successive CR strikes ($\tau_{\rm heat}$). A numerical formula for CRD was presented by \citeauthor{Hasegawa93a}\,(\citeyear{Hasegawa93a}; hereafter HH93), and reads for the desorption rate coefficient of species $i$ in the ice as

\begin{equation}\label{k_HH93}
k_{\rm CRD}^{\rm HH93}(i) = f(a,T_{\rm max}) \, k_{\rm therm}(i,T_{\rm max}) \, ,
\end{equation}
where $a$ is the grain radius, $T_{\rm max}$ is the transient maximum temperature that the grain reaches upon a CR impact, $f(a,T_{\rm max})$ is the ratio $\tau_{\rm cool}/\tau_{\rm heat}$, and $k_{\rm therm}(i,T_{\rm max})$ is the thermal desorption rate coefficient of species $i$. This expression is almost ubiquitously adopted in present day gas-grain chemical models. Making several assumptions, HH93 arrived at a constant value $f(a,T_{\rm max}) = 3.16 \times 10^{-19}$. One of the assumptions they made that is particularly relevant to the present work is that the CRs are composed of iron nuclei only and deposit an average energy of 0.4\,MeV upon passing through a grain, which for spherical grains with a radius of 0.1\,$\mu$m gives $T_{\rm max} = 70\,\rm K$. These constant values of $f(a,T_{\rm max}=70\,\rm K)$ and $T_{\rm max}=70\,\rm K$ are applied in all cases in simulations that incorporate the HH93 CRD model, regardless of the actual (time-dependent) ice composition in the simulation.

Expanding on the HH93 CRD model, we presented in \citeauthor{Sipila21}\,(\citeyear{Sipila21}; hereafter S21) a revised description of CRD where the effect of the time-dependent ice abundances on the desorption efficiency is consistently taken into account, that is, the value of $\tau_{\rm cool}$ depends explicitly on the time-dependent ice composition:

\begin{equation}\label{eq:coolingTime}
\tau_{\rm cool} = \frac{E_{\rm th}}{\dot{E}} =  \frac{E_{\rm th}}{\dot{E}_{\rm subl} + \dot{E}_{\rm rad}} \, ,
\end{equation}
where $E_{\rm th}$ and $\dot{E}_{\rm rad}$ represent the deposited energy and energy lost via radiative cooling per unit time for a grain at a temperature $T_{\rm max}$, and

\begin{equation}\label{eq:evaporationRate}
\dot{E}_{\rm subl} = N_{\rm des} \sum_j k_{\rm B} \, E_b(j) \, \theta(j) \, k_{\rm therm}(j,T_{\rm max})
\end{equation}
is the energy lost via sublimation per unit time. The summation is over the species participating in the cooling (see S21), $N_{\rm des}$ is the number of desorption sites on the grain surface, $k_{\rm B}$ is the Boltzmann constant, while $E_b(j)$ and $\theta(j)$ are respectively the fractional abundance and binding energy of the cooling species on the grain. We also introduced the constraint that the CRD rate coefficient must have an upper limit equal to the grain heating frequency (the inverse of $\tau_{\rm heat}$), which ensures that the desorption events cannot occur more often than the grains are actually struck by CRs:

\begin{equation}\label{eq:k_S21}
k_{\rm CRD}(i) = \min \left( f(a,T_{\rm max}) \, k_{\rm therm}(i,T_{\rm max}) , \tau_{\rm heat}^{-1} \right) .
\end{equation}

The models described above assume that the tracks of CR strikes always go through the center of the grain, that is, the passing CR always deposits the highest possible amount of energy upon impact with the grain. In reality, one expects that CR strikes will often occur at non-zero impact parameters. Paired with the well established fact that CRs do not come in a single flavor associated with a particular energy but instead there is a spectrum of CRs consisting of a variety of species and energies, a distribution of deposited energies and associated values of $T_{\rm max}$, instead of just one fixed value for both, is expected. In particular, collisions that deposit a small amount of energy occur at a much higher rate than collisions that deposit a large amount of energy, meaning that models that consider weak heating events could be expected to predict enhanced desorption of weakly bound species, such as $\rm H_2$.

In the present paper we further expand on the CRD models introduced earlier and consider the effect of the distribution of $T_{\rm max}$ values on the abundances of major interstellar molecules, in particular $\rm H_2$, in comparison to the previous models that assume a constant value $T_{\rm max} = 70\,\rm K$. The $T_{\rm max}$ distribution is treated numerically by transforming Eq.\,(\ref{eq:k_S21}) into a sum over the $T_{\rm max}$ values:

\begin{equation}\label{eq:k_24}
k_{\rm CRD}(i) = \sum_{j} \min \left( f(a,T_{\rm max}^j) \, k_{\rm therm}(i,T_{\rm max}^j) , (\tau_{\rm heat}^j)^{-1} \right) ,
\end{equation}
where the sum runs over the members of the temperature spectrum, that is, the unique values of $T_{\rm max}$ values that we consider. The grain cooling time is now also a function of $T_{\rm max}$ (Eq.\,\ref{eq:coolingTime}) because the energy required to heat the grain to a given temperature increases with $T_{\rm max}$ (see also Sect.\,\ref{ss:input}). We note that, strictly speaking, Eq.\,(\ref{eq:k_24}) should be expressed as an integral because the $T_{\rm max}$ distribution is continuous; we present the discretized form for the purposes of straightforward application in gas-grain chemical models (see also additional discussion below). 

\begin{table*}
	\centering
	\caption{Physical conditions considered in the chemical simulations.}
	\begin{tabular}{cccccc}
		\hline
		\hline
		Label & Environment & Volume density $n({\rm H_2})$\,[cm$^{-3}$] & $T_{\rm gas} = T_{\rm dust}$\,[K] & $A_{\rm V}$\,[mag] & Ice mantle thickness$^a$\,[$\mu$m]\\
		\hline
		E1 & Outer molecular cloud & $10^3$ & 15 & 1.5 & 0.000 \\
		E2 & Inner molecular cloud & $10^4$ & 10 & 3 & 0.010 \\
		E3 & Dense core & $10^5$ & 10 & 9 & 0.025 \\
		\hline
	\end{tabular}
	\tablenotetext{a}{The tabulated mantle thickness corresponds to the value assumed in the calculations of KK22.}
	\label{tab:physicalParameters}
\end{table*}

\subsection{Input data}\label{ss:input}

We adopt the input data for heating frequencies and deposited energies as a function of $T_{\rm max}$ from \citet{Kalvans22b}, who calculated these quantities using CR spectra from \citet{Padovani18} (hereafter P18; their ``Low'' and ``High'' models). Their calculations take into account the possibility of nonzero impact parameter and the attenuation of the CR spectrum as a function of hydrogen column density. They provided tabulated values for a variety of cases involving different grain sizes, ice mantle thicknesses and levels of attenuation.

Here we adopt spherical and monodisperse grains with a radius of 0.1\,$\mu$m to remain consistent with our previous work (S21). We aim to investigate the effect of weak heating events over a variety of physical conditions found across molecular clouds in the interstellar medium, from the low-density outer regions to dense cores within. To achieve this, we associate the levels of attenuation considered by \citet{Kalvans22b} with sets of physical conditions that we consider to be well representative of environments at those extinction values. These are given in Table~\ref{tab:physicalParameters}. \citet{Kalvans22b} considered for each grain size and attenuation level several possibilities for the ice mantle thickness. Here we assume that ice mantles grow thicker with increasing extinction, and we have chosen from the \citet{Kalvans22b} data the thicknesses that are in our view appropriate for each extinction value, as also shown in Table~\ref{tab:physicalParameters}.

We note that the choice of the temperature in each set of physical conditions will affect the abundances predicted by the simulations, but in the present case we concentrate on the effect that weak heating events have on the chemistry as a whole and the absolute abundances are not the primary concern. We do not expect the general conclusions drawn below to change if, for example, we used different temperatures than those given in Table~\ref{tab:physicalParameters} -- as long as the temperature is not high enough for $\rm H_2$ to desorb thermally, which would prevent us from quantifying the effect of the weak heating events (see also Sect.\,\ref{ss:S21vsTmax}).

\begin{figure*}
\centering
        \includegraphics[width=2.0\columnwidth]{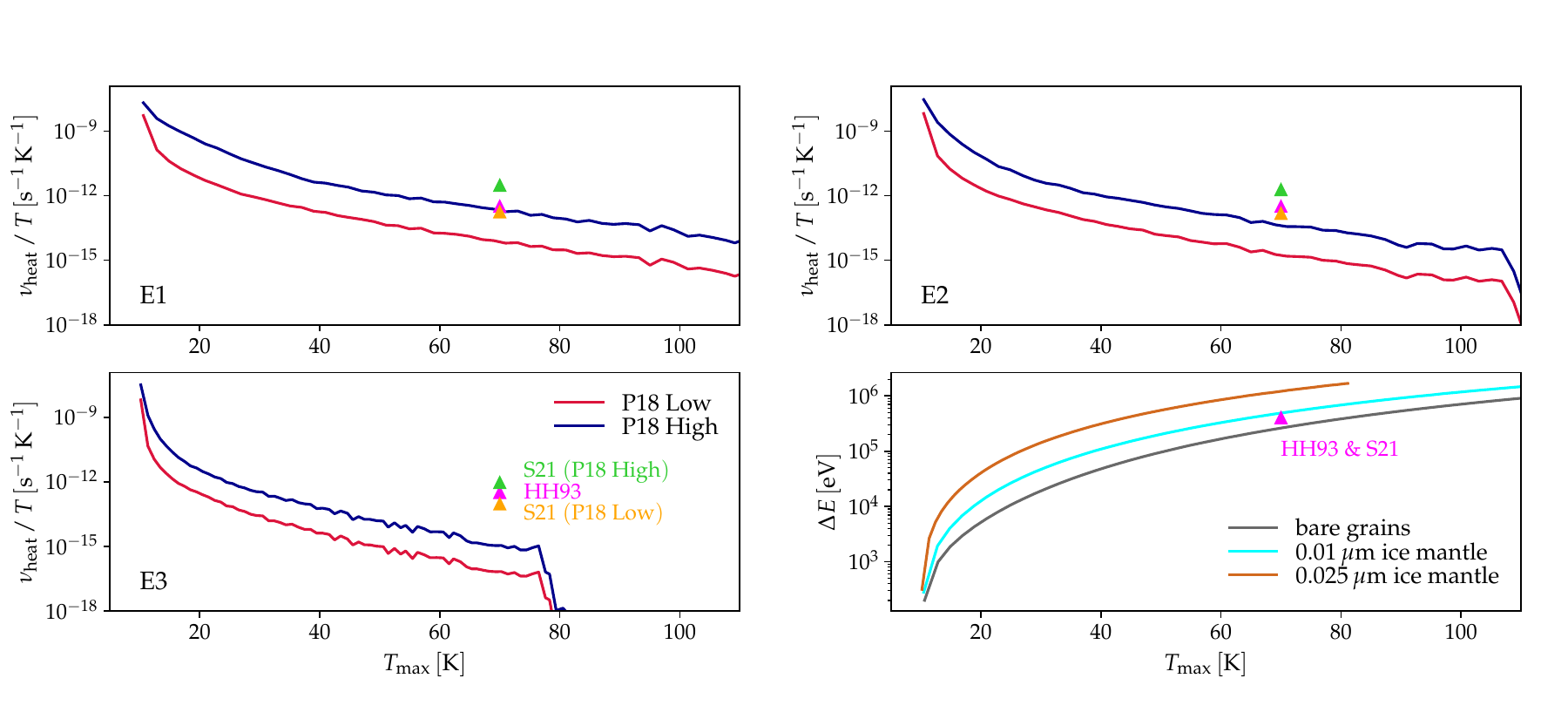}
    \caption{Heating frequencies per unit temperature ($\nu_{\rm heat}/T$) and deposited energies ($\Delta E$) as a function of $T_{\rm max}$ for 0.1\,$\mu$m bare grains (simulation E1, top left), 0.1\,$\mu$m grains coated with a 0.01\,$\mu$m ice mantle (E2, top right), or 0.1\,$\mu$m grains coated with a 0.025\,$\mu$m ice mantle (E3, bottom left). Solid lines reproduce the data from \citet{Kalvans22b} assuming either the P18 Low (red) or High (blue) CR spectrum. Pink, orange, and green triangles indicate the values used by HH93 and S21 for the P18 Low and High CR spectra, respectively. We note that for each value of ice mantle thickness there is only one curve for the deposited energies, because only the heating frequency changes when the CR model is switched. The deposited energy is always constant at 0.4\,MeV in both HH93 and S21.}
    \label{fig:heatingFrequencies}
\end{figure*}

Figure~\ref{fig:heatingFrequencies} shows the \citet{Kalvans22b} data for the heating frequencies and deposited energies as a function of $T_{\rm max}$ for 0.1\,$\mu$m grains for the three pairs of visual extinction and ice mantle thickness given in Table~\ref{tab:physicalParameters}. It is evident that weak heating events occur orders of magnitude more frequently than strong ones, and that the strikes heating a grain up to a given value of $T_{\rm max}$ occur with a rapidly decreasing frequency as the extinction increases.

The predictions of our simulations are sensitive to the values of heating frequencies and deposited energies, and hence to evaluate the robustness of the predictions of \citet{Kalvans22b} on the heating rate and deposited energy, we have performed similar calculations of our own, which are described in Appendix~\ref{a:heatingFrequencyDerivation}; we find very similar values to \citet{Kalvans22b}, implying that the uncertainty on the side of the input values is small. The data of \citet{Kalvans22b} is binned to a series of $T_{\rm max}$ values at a resolution of 1 or 2\,K depending on the value of extinction, with the data spread over at least 50 bins in each case. We have examined the sensitivity of our chemical simulations to the input resolution by rebinning the \citet{Kalvans22b} data to coarser resolution. We found that simulations using a resolution of 10 $T_{\rm max}$ bins still yielded abundances very close (within ten per cent) to those calculated using full resolution input data. The results are thus quite insensitive to the input resolution as long as the bin count is not very small. However, the computational time benefit from using low resolution is also very small, and we always adopt the full resolution data as presented by \citet{Kalvans22b}.

\subsection{Treatment of $\rm H_2$}\label{ss:h2}

Introducing weak heating events into the CRD model leads to the {\sl possibility} of enhanced desorption of lightly bound species. A particularly interesting case in this context is molecular hydrogen given the challenges, outlined above, related to it on the side of observations, laboratory experiments as well as chemical models. CRs heat the grains to temperatures of a few tens of K relatively often, and since these temperatures are not enough for the efficient desorption of CO for example, the cooling of the grain is due to the most lightly bound species, in practice $\rm H_2$ as it is by far the most abundant one. However, if the timescale of $\rm H_2$ adsorption is shorter than the timescale for weak heating events, $\rm H_2$ will still gradually accumulate on the grains despite the frequent CR strikes.

In S21, we had to exclude $\rm H_2$ from the list of cooling molecules because its abundance on the grains was very high in the simulations. Letting $\rm H_2$ participate in the cooling would have meant that virtually all of the energy deposited in a CR strike would have been shed very quickly via the desorption of $\rm H_2$, translating to a short grain cooling time. As the CRD rate coefficients are proportional to the cooling time (Eq.\,\ref{k_HH93}), this meant that there was only a very low probability for species other than $\rm H_2$ to be desorbed while the grain is at a high temperature, leading to extreme overall levels of depletion.

In the present paper, we examine the effect of weak heating events on the population of $\rm H_2$ in the ice using a step by step approach where we sequentially introduce new processes to the model. The starting point is a comparison between S21 and the new CRD model presented here. We then examine cases where the binding energy of $\rm H_2$ is modified depending on its coverage on the grain \citep[following][]{Taquet14}. Finally, we allow for the possibility of $\rm H_2$ cooling the grain and examine how the results of the preceding steps change due to this modification.

\subsection{Chemical model}

The chemical simulations have been performed with our gas-grain chemical code {\sl pyRate}, whose basic properties are described in \citet{Sipila15a}, and which was also used to produce the results presented in S21. Briefly, {\sl pyRate} is a gas-grain chemical code which solves a system of rate equations and outputs time-dependent number densities for the species included in the simulation. Gas-phase and grain-surface chemistry are connected via adsorption and several desorption mechanisms (thermal desorption, photodesorption, chemical desorption, CRD). We use here the same chemical networks as in S21 (with deuterium and spin-state chemistry), which are based on the KIDA 2014 public release \citep{Wakelam15} and contain over 70,000 reactions in the gas phase and approximately 2000 reactions on grain surfaces. The model is versatile and can be applied to a range of environments, from low to high densities as we do here. The simulations are inherently zero-dimensional (though they can be coupled with a physical model, as in S21 for example), and indeed in this paper we calculate a set of simple zero-dimensional models adopting the physical conditions given in Table~\ref{tab:physicalParameters}.

\begin{table}
	\centering
	\caption{Initial abundances (with respect to the total hydrogen number density $n_{\rm H}$) used in the chemical simulations.}
	\begin{tabular}{cc}
		\hline
		\hline
		Species & Abundance\\
		\hline
		$\rm H_2$$^{a}$ & $5.00\times10^{-1}$\\
		$\rm He$ & $9.00\times10^{-2}$\\
		$\rm HD$ & $1.60\times10^{-5}$\\
		$\rm C^+$ & $1.20\times10^{-4}$\\
		$\rm N$ & $7.60\times10^{-5}$\\
		$\rm O$ & $2.56\times10^{-4}$\\
		$\rm S^+$ & $8.00\times10^{-8}$\\
		$\rm Si^+$ & $8.00\times10^{-9}$\\
		$\rm Na^+$ & $2.00\times10^{-9}$\\
		$\rm Mg^+$ & $7.00\times10^{-9}$\\
		$\rm Fe^+$ & $3.00\times10^{-9}$\\
		$\rm P^+$ & $2.00\times10^{-10}$\\
		$\rm Cl^+$ & $1.00\times10^{-9}$\\
		\hline
	\end{tabular}
	\label{tab:initialAbundances}
	\tablenotetext{a}{The initial $\rm H_2$ ortho/para ratio is $1.0 \times 10^{-3}$.}
\end{table}

For this work, pyRate was updated so that a distribution of $T_{\rm max}$ values and deposited energies can be taken into account. We use the same initial abundances as in S21 (originating in Table~1 in \citealt{Sipila19a}), reproduced here in Table~\ref{tab:initialAbundances}. We always adopt a cosmic ray ionization rate $\zeta = 1.3 \times 10^{-17} \, \rm s^{-1}$ in connection with the P18 Low model, and $\zeta = 1.0 \times 10^{-16} \, \rm s^{-1}$ in connection with the P18 High model. These constant values are chosen for the sake of simplicity, noting that, in a realistic case, due to attenuation one does not expect a single ionization rate to apply across a range of physical conditions. The dust-to-gas mass ratio is set to 0.01, the UV field external to the core is taken from \citet{Black94}, and $\rm H_2$ self-shielding is simulated using the parameterized approach of \citet{Draine96}. We concentrate the present analysis on 2-phase chemical models (where the ice on the grains is treated as a single reactive layer), noting that the method can be applied just as well in 3-phase models (ice separated into a reactive layer and a possibly reactive bulk beneath).

\section{Results}\label{s:results}

The next subsections present the results of our simulations. We adopt the P18 Low CR spectrum for all simulations discussed in the main text; the recent studies by \citet{Obolentseva24} and \citet{Neufeld24} strongly favor low CR ionization rates in accordance with the Low model. Additional results for the P18 High CR spectrum are given in Appendix~\ref{a:P18H}.

\subsection{S21 vs. variable $T_{\rm max}$ CRD}\label{ss:S21vsTmax}

\begin{figure*}
\centering
        \includegraphics[width=2.0\columnwidth]{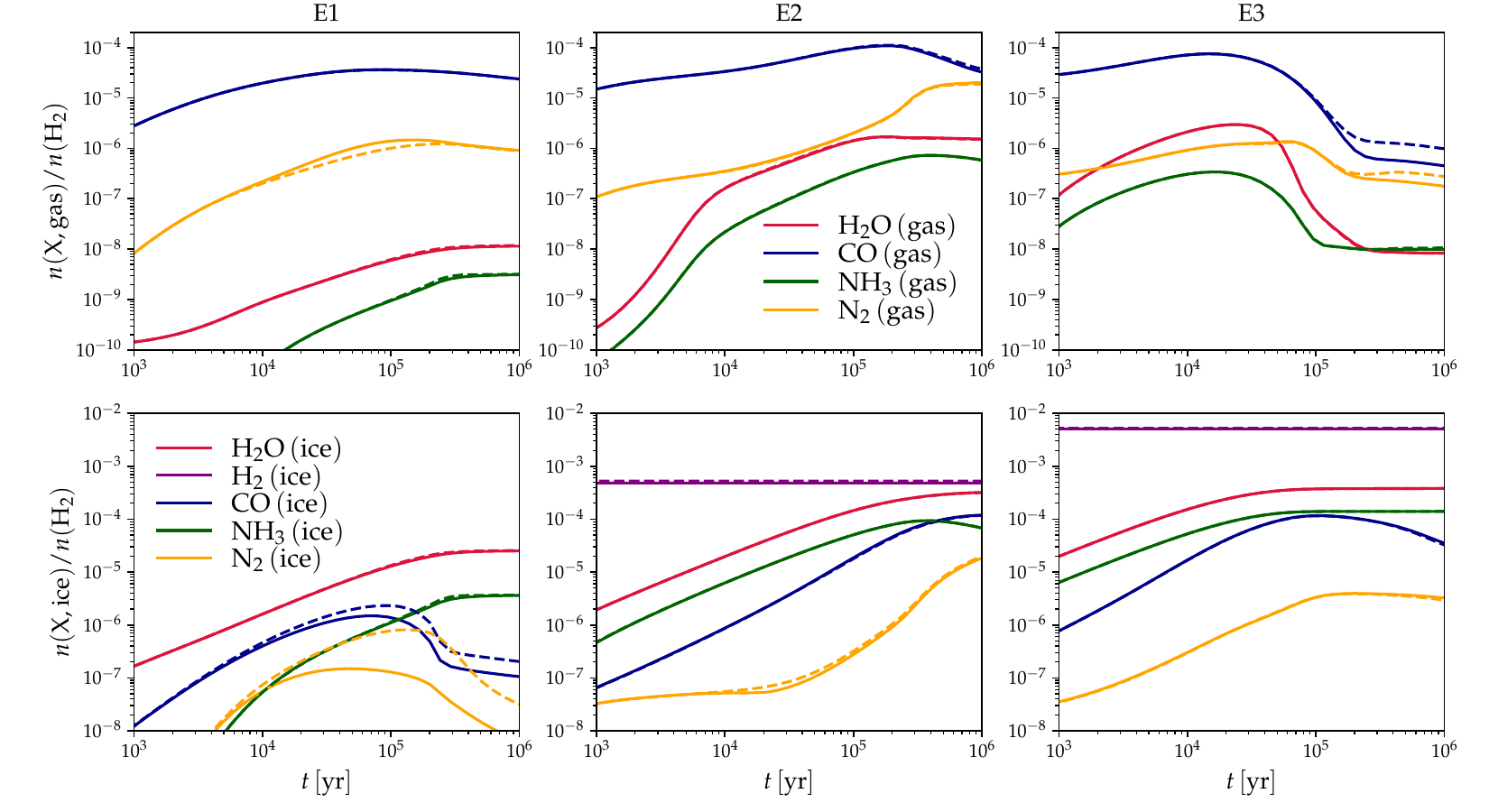}
    \caption{Abundances with respect to $\rm H_2$ of selected gas phase (top) and grain-surface (bottom; labeled as ``ice'') species as a function of time in physical conditions E1 to E3 (columns from left to right), assuming the P18 Low CR spectrum. Solid lines indicate results of simulations using the variable $T_{\rm max}$ CRD model, while dashed lines indicate results of simulations using the S21 CRD model.}
    \label{fig:abundances}
\end{figure*}

Figure~\ref{fig:abundances} displays simulated abundances for common gas-phase and ice species, for the three sets of physical conditions tabulated in Table~\ref{tab:physicalParameters}, in the gas phase and in the ices on grain surfaces calculated using our fiducial model in which $\rm H_2$ is excluded from the list of grain coolants (as was done in S21). There is overall very little difference between the simulation results, and in many cases the two sets of curves overlap almost perfectly. Small differences in the CO and $\rm N_2$ gas-phase abundances occur at high density, and in solid abundances for the same species at low density.

Let us examine first the E1 set of physical conditions. CO and $\rm N_2$, both relatively lightly bound to the grain surface, are desorbed more efficiently in the variable $T_{\rm max}$ model as compared to S21, which is due to light CR species striking the grains often. The effect is larger for $\rm N_2$ than for CO because of their binding energies; our model assumes $E_{\rm b}({\rm N_2}) = 1000\,\rm K$ and $E_{\rm b}({\rm CO}) = 1150\,\rm K$ \citep{Garrod06}. The increased desorption efficiency has only a limited effect on the $\rm N_2$ and CO gas-phase abundances, however, owing to the gas-phase species being much more abundant -- this is especially clear in the case of CO. The weak heating events affect the $\rm H_2O$ and $\rm NH_3$ abundances in the ice only very marginally; they are both strongly bound to the grains, with binding energies of several thousand Kelvin, and hence their desorption rates are low. The temperature (15\,K) is high enough for hydrogen to be thermally desorbed, and hence $\rm H_2$ is not visible on the plot.

\begin{figure}
\centering
        \includegraphics[width=1.0\columnwidth]{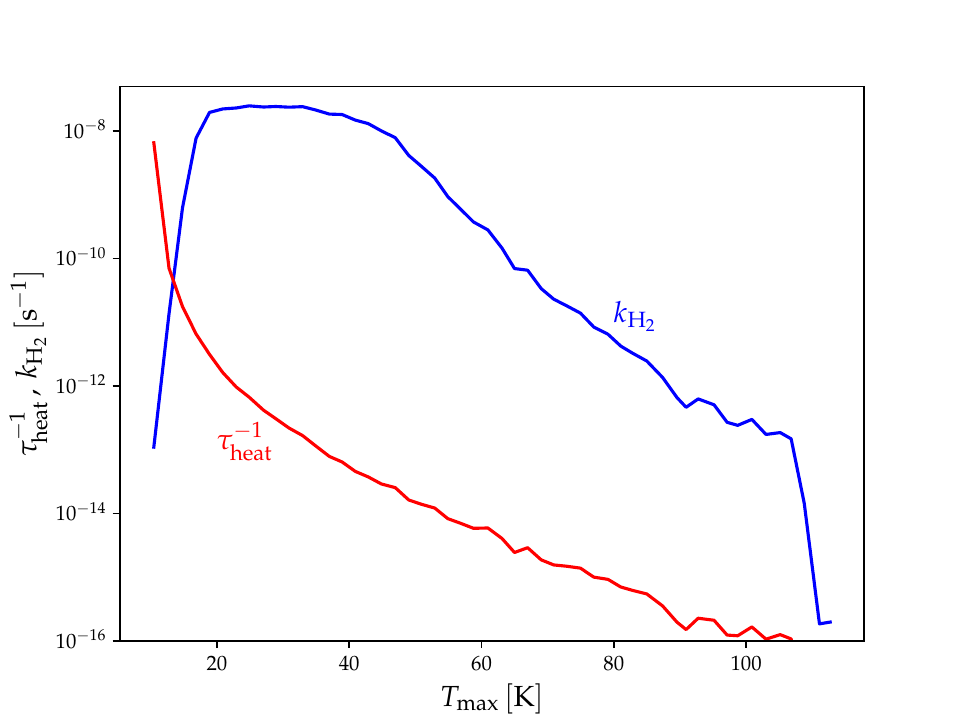}
    \caption{Nonlimited CRD rate coefficient of $\rm H_2$ ($k_{\rm H_2}$; see text) and the grain heating frequency (inverse of the heating time; $\tau_{\rm heat}^{-1}$) as a function of $T_{\rm max}$ in the E2 model. The evolutionary time, which affects the grain cooling time and hence the value of $f(0.1\mu{\rm m},T_{\rm max})$ at each $T_{\rm max}$, is set to $t = 10^5 \, \rm yr$.}
    \label{fig:H2_CRD_rate}
\end{figure}

Moving to the E2 set of physical conditions, the higher volume density and lower temperature (10\,K) cause a significant amount of $\rm H_2$ to be adsorbed on the grains, and the $\rm H_2$ ice abundance is very high in the present model as well as in S21. There is overall very little difference between the two sets of simulation results; as noted above, we expect potential effects only for the most lightly bound species, but even for $\rm H_2$ there is almost no difference in the abundance predicted by the two simulations. To understand this, let us examine in detail the efficiency of CRD for $\rm H_2$ for different values of $T_{\rm max}$. In our model, the CRD rate coefficients are limited by the frequency of CR heating events (Eq.\,\ref{eq:k_S21}), which corresponds to the physical requirement that any one molecule on the grain surface should not be able to be desorbed due to CRD more often than the grain is actually struck by a CR (the strike frequency is given by $\tau_{\rm heat}^{-1}$) -- without the limiting term in the rate coefficient, such a situation can arise for weakly bound species like $\rm H_2$. Figure~\ref{fig:H2_CRD_rate} shows the dependence of $k_{\rm H_2} \equiv f(0.1\mu{\rm m},T_{\rm max})\,k_{\rm therm}({\rm H_2},T_{\rm max})$ and $\tau_{\rm heat}^{-1}$ on $T_{\rm max}$ in the E2 model. Evidently, $k_{\rm H_2}$ is high for $T_{\rm max} \sim 20-50\,\rm K$, and based on this one might expect a significant amount of $\rm H_2$ to be desorbed due to the weak heating events. However, for $T_{\rm max}$ greater than approximately 14\,K, $k_{\rm H_2}$ always (greatly) exceeds $\tau_{\rm heat}^{-1}$ and hence $k_{\rm CRD}({\rm H_2}) = min(k_{\rm H_2},\tau_{\rm heat}^{-1}) = \tau_{\rm heat}^{-1}$ in the entire range $T_{\rm max} > 14\,\rm K$. CR strikes heating the grain to $T_{\rm max} < 14\,\rm K$ occur often, but these temperatures are too low to induce efficient $\rm H_2$ desorption. The total $\rm H_2$ CRD rate coefficient is a sum over the $T_{\rm max}$ distribution (Eq.\,\ref{eq:k_24}) and hence the actual value of $k_{\rm CRD}({\rm H_2})$ cannot be simply read off Fig.\,\ref{fig:H2_CRD_rate} (see also below), but nevertheless the conclusion is that the weak heating events do not actually lead to an overall increase of $\rm H_2$ desorption because the desorption is limited by the frequency of CR heating. Increasing the CR flux does however lead to a small increase in desorption (see Appendix~\ref{a:P18H}). We also note that the crossing point of the $k_{\rm H_2}$ and $\tau_{\rm heat}^{-1}$ curves is of course affected by the assumption on the binding energy of $\rm H_2$; we discuss lower values of the binding energy in Sect.\,\ref{ss:dynamicEb}.

\begin{figure}
\centering
        \includegraphics[width=0.8\columnwidth]{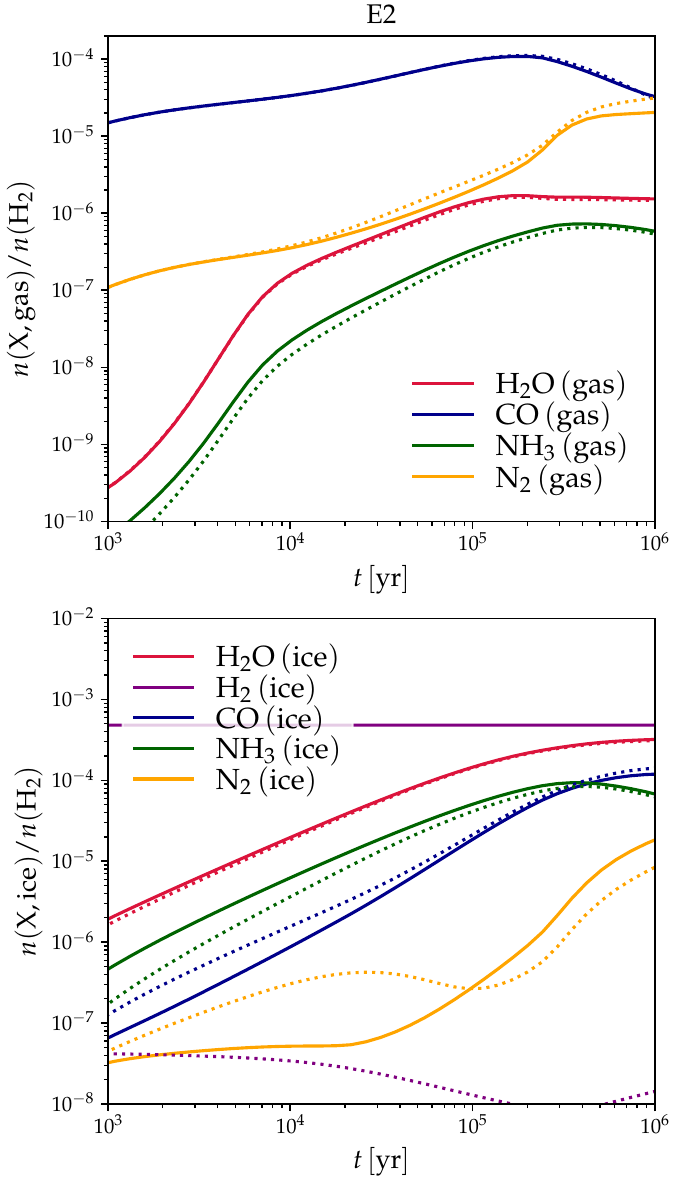}
    \caption{Abundances with respect to $\rm H_2$ of selected gas phase (top) and grain-surface (bottom; labeled as ``ice'') species as a function of time in physical conditions E2, assuming the P18 Low CR spectrum. Solid lines indicate results of simulations using the variable $T_{\rm max}$ CRD model, while dotted lines indicate results of simulations where the limiting factor in Eq.\,\ref{eq:k_24} is disabled.}
    \label{fig:rateLimit_vs_noRateLimit}
\end{figure}

If the rate coefficient limitation is (unphysically) removed, the simulation results are decidedly different for $\rm H_2$. This is demonstrated in Fig.\,\ref{fig:rateLimit_vs_noRateLimit}, which shows the results of a simulation carried out with the variable $T_{\rm max}$ CRD approach but where the limiting factor in Eq.\,\ref{eq:k_24} is disabled. This change leads to a drastic reduction in the abundance of $\rm H_2$ in the ice, as well as (minor) modifications to the abundances of other species that are somewhat more strongly bound than $\rm H_2$, for example $\rm N_2$. We emphasize that these solutions are unphysical. The necessity of including the limiting term in the CRD rate coefficient is clear, as without it the CRD rate coefficients of lightly-bound species can grow to very large values that greatly exceed the physical constraint that desorption events cannot occur more often than the grains are actually heated by CRs.

\begin{figure*}
\centering
        \includegraphics[width=2.0\columnwidth]{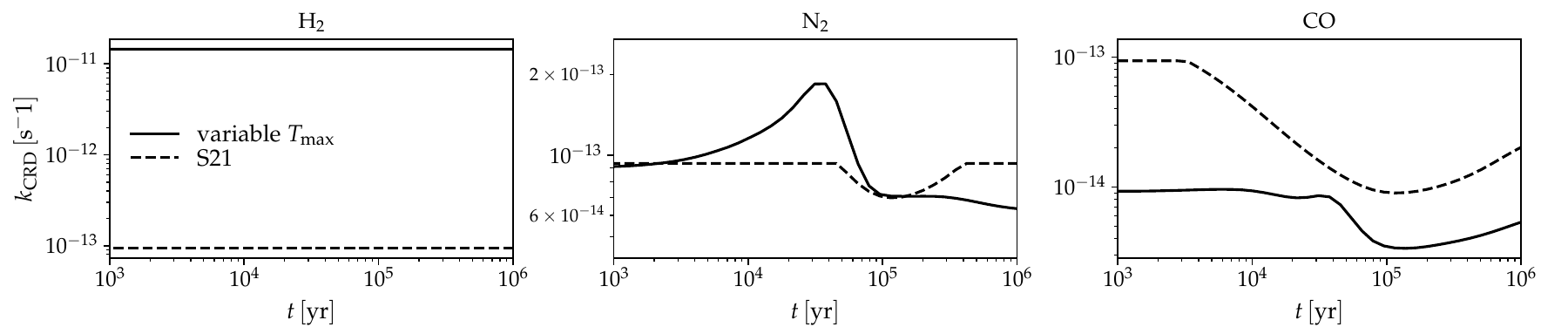}
    \caption{Time-dependent desorption rate coefficients ($k_{\rm CRD}$) of $\rm H_2$, $\rm N_2$, and CO in physical conditions~E3. Solid lines show results calculated with the variable $T_{\rm max}$ model, while dashed lines show results calculated using the S21 CRD model.}
    \label{fig:rateCoefficients}
\end{figure*}

At the highest density probed by our simulations (physical conditions E3), there is again very little difference in terms of abundances arising from the introduction of the weak heating events. There is small variation in the CO and $\rm N_2$ gas-phase abundances between the models at late times as a result of the complex interplay between time-dependent grain cooling times and $T_{\rm max}$-dependent CRD efficiencies. To support the interpretation of the predicted abundance curves, we show in Fig.\,\ref{fig:rateCoefficients} the time-dependent CRD rate coefficients ($k_{\rm CRD}$) of $\rm H_2$, $\rm N_2$, and CO. It is evident that the CO (binding energy 1150\,K) desorption efficiency is at late times higher in the S21 model, and consequently CO depletes slightly more in the variable $T_{\rm max}$ model (Fig.\,\ref{fig:abundances}). For $\rm N_2$ (binding energy 1000\,K) there is a crossover point at $t \sim 10^5$\,yr where the CRD rate in the variable $T_{\rm max}$ model drops below that of S21, and this flip is traced by the abundance curves in Fig.\,\ref{fig:abundances}, although at times earlier than $t \sim 10^5$\,yr the difference in the two orange curves is so small so as to be only very slightly visible in the plot; this is because of the low amount of $\rm N_2$ in the ice at early times in the simulation, translating to a low desorption rate.

In summary of the above, the introduction of weak heating events influences gas-phase and ice abundances across a variety of physical conditions only to a small degree, and for the most part the results of the variable $T_{\rm max}$ models are almost identical to those of S21, with the former model predicting in particular only a very minor increase in $\rm H_2$ desorption. The small difference is due to the requirement that CRD events cannot occur more often than the grains are actually heated by CRs. In the specific case of $\rm H_2$, it is evident that even when taking weak heating events into account, adsorption always overcomes (CR) desorption, and other ways of regulating the $\rm H_2$ ice abundance remain necessary.

\subsection{Impact of CRD on dynamic $\rm H_2$ binding energy models}\label{ss:dynamicEb}

\begin{figure}
\centering
        \includegraphics[width=0.8\columnwidth]{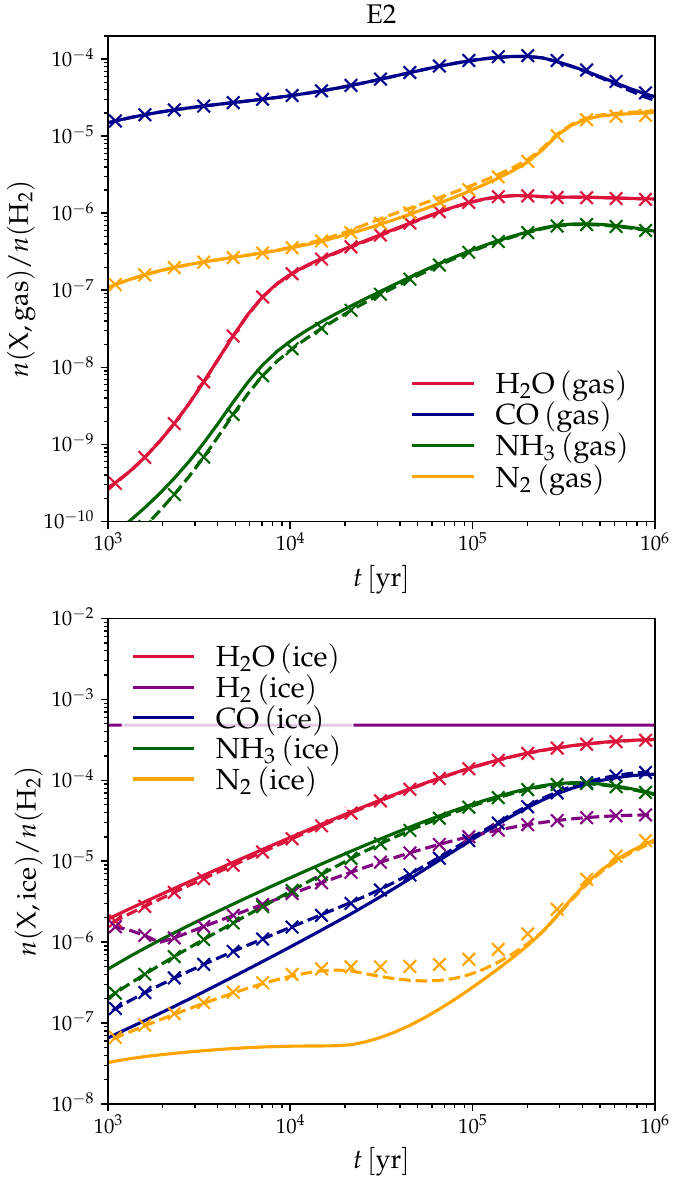}
    \caption{Abundances with respect to $\rm H_2$ of selected gas phase (top) and grain-surface (bottom; labeled as ``ice'') species as a function of time in physical conditions E2, assuming the P18 Low CR spectrum. Solid lines indicate results of simulations using the variable $T_{\rm max}$ CRD model, while dashed lines show the results of an otherwise identical simulation but where the binding energy of $\rm H_2$ varies with time according to Eq.\,\ref{eq:H2bindingEnergy}. As a reference, we also show the results of the corresponding S21 model where the time-dependent $\rm H_2$ binding energy is applied (crosses).}
    \label{fig:normal_vs_dynamicEb}
\end{figure}

We compare here the results of our variable $T_{\rm max}$ CRD model to one where the binding energy of $\rm H_2$ varies as a function of the $\rm H_2$ coverage in the ice, based on the idea that the binding energy of $\rm H_2$ on $\rm H_2$ is much lower than on water. For this we use a slightly modified version of the formulation of \citet{Taquet14}:

\begin{align}\label{eq:H2bindingEnergy}
E_{\rm b}({\rm H_2}) &= (1 - \theta({\rm H_2}) - \theta({\rm bare}))\,E_b^{\rm H_2O}({\rm H_2}) \, +\\
				& \theta({\rm H_2})\,E_b^{\rm H_2}({\rm H_2}) + \theta({\rm bare})\,E_b^{\rm bare}({\rm H_2}) \, , \nonumber
\end{align}
where $\theta({\rm H_2O})$ and $\theta({\rm bare})$ represent respectively the coverages of $\rm H_2$ and unoccupied binding sites on the  grain surface, while $E_b^{\rm H_2O}({\rm H_2})$ and $E_b^{\rm bare}({\rm H_2})$ represent respectively the binding energy of $\rm H_2$ on water ice and on bare grains (taken from Table~2 in \citealt{Taquet14}). The binding energy of $\rm H_2$ on $\rm H_2$ is calculated as 

\begin{equation}
E_b^{\rm H_2}({\rm H_2}) = E_b^{\rm H_2}({\rm H}) \times \frac{E_b^{\rm H_2O}({\rm H_2})}{E_b^{\rm H_2O}({\rm H})} \, ,
\end{equation}
where the binding energy of H on $\rm H_2$ is 45\,K \citep{Vidali91}. The formulation above differs from \citet{Taquet14} in that the effect of pure ices is not taken into account, and from that of \citet{Garrod11} in that bare grain surfaces are accounted for. The latter is, however, important only at relatively low volume densities and early evolutionary times when little ice has accumulated on the grains.

Figure~\ref{fig:normal_vs_dynamicEb} shows the results of simulations where the $\rm H_2$ binding energy is updated time-dependently based on Eq.\,\ref{eq:H2bindingEnergy}. Applying the time-dependent $\rm H_2$ binding energy decreases the abundance of $\rm H_2$ in the ice by over an order of magnitude as compared to the fiducial simulation. This is in the present model due to two effects. Firstly, the $\rm H_2$ binding energy decreases from its fiducial value down to a minimum of approximately 420\,K during the course of the simulation, which translates to an increase of the desorption rate coefficient by a factor of $\sim$3 (for $T_{\rm max} = 70 \, \rm K$) compared to the model with constant $\rm H_2$ binding energy (500\,K). Secondly, even though $\rm H_2$ itself does not participate in the cooling of the grain following a CR strike, the reduction in the $\rm H_2$ abundance in the ice affects the cooling time indirectly because the relative populations of the coolants (see Eq.\,\ref{eq:evaporationRate}) are calculated with respect to the total ice abundance including $\rm H_2$. In practice the grain cooling time decreases in the time-dependent $\rm H_2$ binding energy model, as compared to the fiducial one, by a factor of 2-3 depending on the simulation time. Hence the total combined effect on the $\rm H_2$ desorption rate can thus be on the order of magnitude level. Disregarding the evolution of grain-surface $\rm H_2$, the simulation results are overall almost unchanged\footnote{The early-time effects seen for example in the case of $\rm N_2$ are due to the variations in the grain cooling time.} when switching between the S21 and the variable $T_{\rm max}$ CRD schemes. This demonstrates clearly that the effect of weak heating events on the abundance of $\rm H_2$ in the ice is very small regardless of the size of the $\rm H_2$ population on the grains -- this result is accordance with what was discussed above regarding the constraints on the frequency of CR strikes.

\subsection{Effect of $\rm H_2$ on the cooling of grains}

\begin{figure*}
\centering
        \includegraphics[width=2.0\columnwidth]{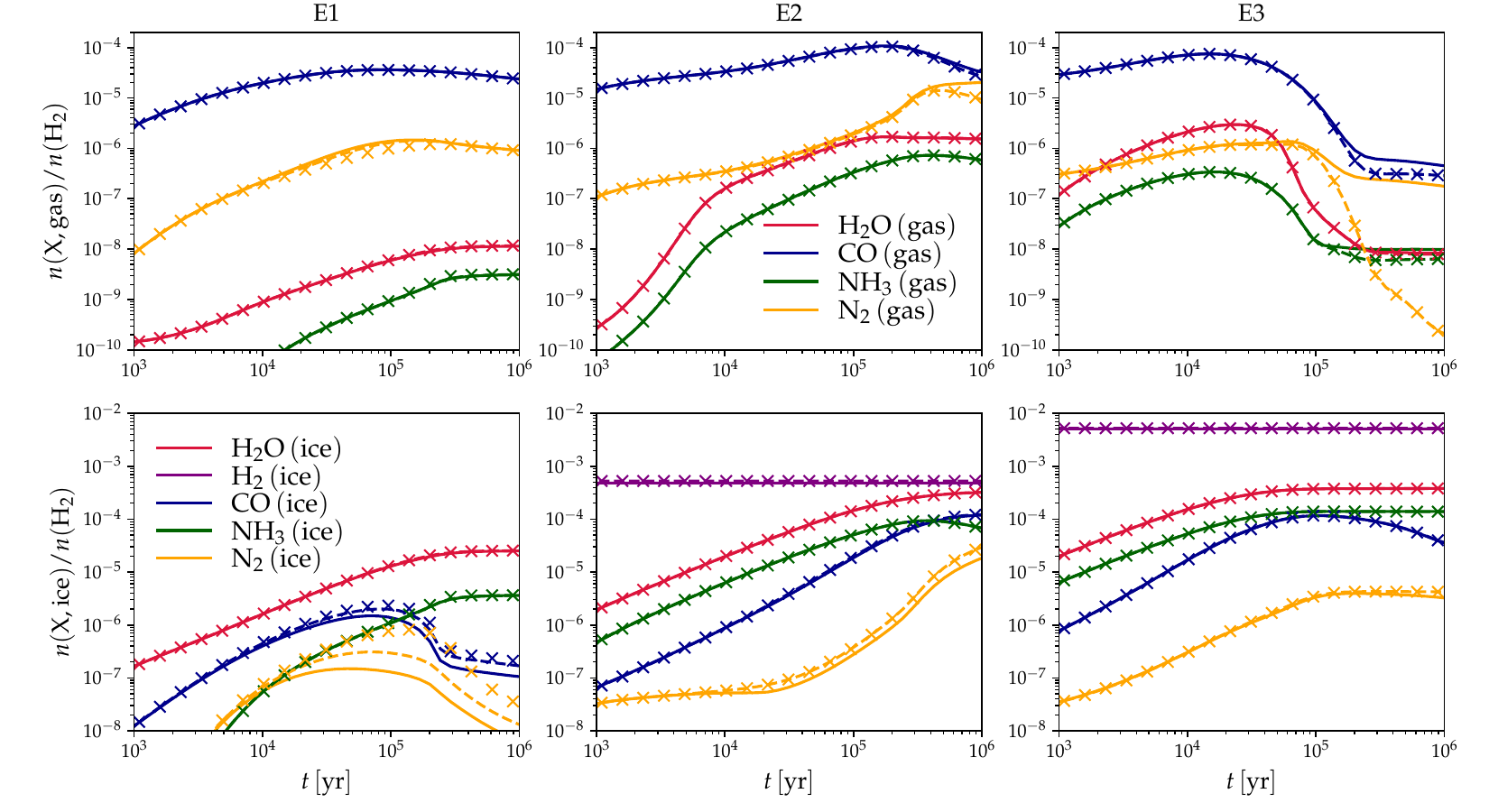}
    \caption{Abundances with respect to $\rm H_2$ of selected gas phase (top) and grain-surface (bottom; labeled as ``ice'') species as a function of time in physical conditions E1 to E3 (columns from left to right), assuming the P18 Low CR spectrum. Solid lines indicate results of simulations using the variable $T_{\rm max}$ CRD model, while dashed lines show the results of an otherwise identical simulation but where $\rm H_2$ desorption is allowed to cool the grain. Crosses indicate the results of the S21 model where the $\rm H_2$ cooling is applied.}
    \label{fig:normal_vs_H2Cooling}
\end{figure*}

$\rm H_2$ is one of the major constituents of the ice on dust grains, and being very lightly bound, it is also one of the most prominent agents for cooling the grains following a CR heating event. If the abundance of $\rm H_2$ in the ice was extremely high, as chemical models without a bespoke treatment of $\rm H_2$ suggest, the majority of the energy deposited by a CR would be dissipated via evaporation of $\rm H_2$ which is expected to lead to a short grain cooling time and hence to very little desorption of species other than $\rm H_2$. As already mentioned in Sect.\,\ref{ss:h2}, it is for this reason that we did not include $\rm H_2$ in the list of coolants in S21. It is however prudent to re-examine that assumption here in the context of the newly introduced weak heating events.

Figure~\ref{fig:normal_vs_H2Cooling} displays the effect that excluding or including $\rm H_2$ included in the list of cooling species has on the simulation results. Firstly, the abundances predicted by the variable $T_{\rm max}$ and S21 models with $\rm H_2$ cooling included are extremely similar, even more so than in the case of no $\rm H_2$ cooling (Fig.\,\ref{fig:abundances}). Secondly, there is a greater degree of depletion at high volume densities in the models with $\rm H_2$ cooling as compared to the fiducial model, which is due to a significant reduction in the grain cooling time; the energy deposited by the CRs is efficiently dissipated via the desorption of a small part of the large $\rm H_2$ reservoir in the ice. The increased depletion is by far the most evident for $\rm N_2$, and we note that the relative increase in the abundance of $\rm N_2$ in the ice in the $\rm H_2$ cooling models compared to the fiducial case is small because the gas-phase reservoir constitutes less than 10\% of all $\rm N_2$ (at late simulation times). The major $\rm N_2$ depletion has a large impact on species derived from it, such as $\rm N_2H^+$ (not shown), which contradicts the observational fact that $\rm N_2H^+$ is found in abundance toward dense cores (see, e.g., \citealt{Punanova18} and references therein). As there is no reason to expect that in a realistic case $\rm H_2$ would not be able to cool the grains just like any other (lightly bound) ice species, our tests support the idea that the amount of $\rm H_2$ ice on interstellar grains cannot be very high. Furthermore, the simulation results clearly indicate that great care must be taken when incorporating $\rm H_2$ as a coolant in chemical models; if its abundance on grains is very high and the resultant grain cooling time is short, the simulation results will not be trustworthy.

\section{Discussion}\label{s:discussion}

The present CRD model makes one fundamental assumption: we consider monodisperse grains with the canonical radius of 0.1\,$\mu$m. The larger the grain, the larger the amount of energy required to heat it to a given value of $T_{\rm max}$. It is conceivable that for smaller grains the heating events where enough energy to efficiently desorb $\rm H_2$ is deposited could occur often enough so as to overpower the continual (re)adsorption, though we note that, in the case of CO, \citet{Silsbee21} have predicted that the desorption timescale has a minimum around 0.1\,$\mu$m (under the assumption of constant ice mantle thickness across the grain size distribution). Whether $\rm H_2$ behaves differently in comparison with CO needs to be checked via dedicated simulations that allow for the possibility of nonconstant ice mantle thickness. Indeed, the new CRD formalism introduced here is straightforward to extend to the case of grain size distributions. The majority of chemical models that include CRD are based on the assumption that CRs transiently heat monodisperse grains to $T_{\rm max} = 70\, \rm K$, and treatments of size distributions have been presented previously by \citet{Iqbal18}, \citet{Zhao18}, \citet{Kalvans19}, and \citet{Sipila20}. In the context of the present work it is also particularly prudent to highlight that \citet{Kalvans20} have studied the contribution of $\rm H_2$ on the cooling of grains. However, none of these previous works have incorporated the full spectrum of peak temperatures into chemical models, as we have done here.

We found that weak heating events do have an effect on species other than $\rm H_2$; differences in simulation results depending on the treatment of CRD arise for ice species at low densities, and for gas phase species at high densities (Fig.\,\ref{fig:abundances}), though in the latter case the variations are small. At low (column) densities, the desorption of weakly bound species such as $\rm N_2$ benefits from frequent weak heating events because the CR flux is not substantially attenuated (Fig.\,\ref{fig:heatingFrequencies}). The resulting decrease in ice abundances (as compared to the classic $T_{\rm max} = 70\,\rm K$ models) could potentially have chemical implications in the context of a collapse model where the gas is allowed to evolve from low to high densities. Quantifying this is however outside the scope of the present work, where we concentrate on constant physical conditions.

Finally, we highlight the great importance of the limiting factor in Eq.\,\ref{eq:k_24}, which ensures that CRD rate coefficients cannot exceed the frequency at which the CR heating events actually occur. We introduced this limit already in S21 and explored here its importance for $\rm H_2$ specifically. Indeed we found that it has a fundamental impact on $\rm H_2$ (and to a lesser extent on other species) as evidenced by Fig.\,\ref{fig:rateLimit_vs_noRateLimit}. Models that do not include this constraint overestimate the CRD rate coefficients of weakly bound species regardless of whether the model considers a single value or multiple values of $T_{\rm max}$, or a constant grain cooling time as opposed to the time-dependent approach that we follow here. For the conventional $T_{\rm max} = 70\,\rm K$ models of HH93 with a heating frequency of $3.16\times10^{-14} \, \rm s^{-1}$ and a grain cooling time of $10^{-5}$\,s, the CRD rate coefficient exceeds the heating frequency for binding energies on the order of 1000\,K, meaning that the CRD rates of all species with binding energy lower than approximately 1000\,K are overstimated. The overestimation grows more severe with decreasing values of the binding energy. We therefore highly recommend the adoption of the limiting factor in all CRD treatments, even those based on the HH93 approach that do not include the advancements presented here and in S21.

\section{Conclusions}\label{s:conclusions}

We have presented a new model of CRD where the grains are allowed to reach a variable maximum temperature upon a CR impact at frequencies that depend on the input CR spectrum, attenuation of the CR flux as a function of column density, and the thickness of the ice mantle on the grains. The new model improves on existing CRD models by taking into account specifically the effect of weak heating events that are able to transiently heat the grains to temperatures of several tens of Kelvin, which is enough to induce desorption of lightly bound species with binding energies of up to, approximately, 1000\,K. Our formalism is easily applicable in rate-equation based gas-grain chemical models, though the CRD rate coefficients depend on the grain heating frequencies and deposited energies associated with the CR strikes, which have to be separately calculated. As in \citet{Sipila21}, we calculated the grain cooling time as function of the evolving ice composition. 

We studied the impact of weak CR heating events on the desorption of $\rm H_2$ in particular and found that, perhaps surprisingly, the $\rm H_2$ ice abundance is almost unaffected by the inclusion of weak heating events in the desorption model. The main reason for this is that the CR strikes do not occur often enough for the additional desorption (as compared to previous models that only consider heavy CRs) to overpower adsorption -- more $\rm H_2$ is adsorbed onto the grains in a given time interval than CRD can remove, despite the weak heating events that occur relatively often. Other ways of keeping the $\rm H_2$ ice population in check remain necessary, such as the time-dependent modification of $\rm H_2$ binding energy based on the amount of $\rm H_2$ on the grain. The inclusion of weak heating events does affect the abundance of species other than $\rm H_2$ in a (column) density-dependent manner, leading to a reduction of the abundances of weakly-bound ice species (e.g., $\rm N_2$) at low densities, and and increase in gas-phase abundances at high densities, though the latter effect is quite minor.

We obtained very low $\rm H_2$ abundances in the ice, also at high densities where the CR flux is strongly attenuated, when we removed the limiting factor in the CRD rate coefficient calculation that prevents the rate coefficients from rising higher than the CR strike frequency. We proposed this factor originally in \citeauthor{Sipila21}\,(\citeyear{Sipila21}; cf. Eq.\,\ref{eq:k_24} in the present paper). Solutions obtained without the limiting factor are however unphysical because the desorption rate coefficients of weakly bound species (like $\rm H_2$) may exceed the rate at which the grains are actually heated by CRs, and the problem is exacerbated by the adoption of weak heating events in the chemical model. This constraint should be applied in all CRD models, even those that follow the classic approach of \citet{Hasegawa93a}.

A major assumption in the present work is that the grains are monodisperse and spherical with a radius of 0.1\,$\mu$m. It is however conceivable that small grains could be heated by CRs often enough that $\rm H_2$ could be efficiently desorbed, and it is therefore of interest to extend the present model to grain size distributions. This will the subject of a future study.

\acknowledgments
The authors thank the anonymous referee for a constructive report that helped to make the presentation of our results clearer. The financial support of the Max Planck Society is acknowledged.

\bibliographystyle{aasjournal}
\bibliography{refs.bib}

\begin{thebibliography}{}
\expandafter\ifx\csname natexlab\endcsname\relax\def\natexlab#1{#1}\fi
\providecommand{\url}[1]{\href{#1}{#1}}
\providecommand{\dodoi}[1]{doi:~\href{http://doi.org/#1}{\nolinkurl{#1}}}
\providecommand{\doeprint}[1]{\href{http://ascl.net/#1}{\nolinkurl{http://ascl.net/#1}}}
\providecommand{\doarXiv}[1]{\href{https://arxiv.org/abs/#1}{\nolinkurl{https://arxiv.org/abs/#1}}}

\bibitem[{{Black}(1994)}]{Black94}
{Black}, J.~H. 1994, in The First Symposium on the Infrared CASP Conf. Ser.,
  58, 355

\bibitem[{{Chuang} {et~al.}(2018){Chuang}, {Fedoseev}, {Qasim}, {Ioppolo}, {van
  Dishoeck}, \& {Linnartz}}]{Chuang18b}
{Chuang}, K.~J., {Fedoseev}, G., {Qasim}, D., {et~al.} 2018, \aap, 617, A87,
  \dodoi{10.1051/0004-6361/201833439}

\bibitem[{{Cuppen} {et~al.}(2009){Cuppen}, {van Dishoeck}, {Herbst}, \&
  {Tielens}}]{Cuppen09}
{Cuppen}, H.~M., {van Dishoeck}, E.~F., {Herbst}, E., \& {Tielens}, A.~G.~G.~M.
  2009, \aap, 508, 275, \dodoi{10.1051/0004-6361/200913119}

\bibitem[{{Draine} \& {Bertoldi}(1996)}]{Draine96}
{Draine}, B.~T., \& {Bertoldi}, F. 1996, \apj, 468, 269, \dodoi{10.1086/177689}

\bibitem[{{Garrod} \& {Herbst}(2006)}]{Garrod06}
{Garrod}, R.~T., \& {Herbst}, E. 2006, \aap, 457, 927,
  \dodoi{10.1051/0004-6361:20065560}

\bibitem[{{Garrod} \& {Pauly}(2011)}]{Garrod11}
{Garrod}, R.~T., \& {Pauly}, T. 2011, \apj, 735, 15,
  \dodoi{10.1088/0004-637X/735/1/15}

\bibitem[{{Hasegawa} \& {Herbst}(1993{\natexlab{a}})}]{Hasegawa93b}
{Hasegawa}, T.~I., \& {Herbst}, E. 1993{\natexlab{a}}, \mnras, 263, 589

\bibitem[{{Hasegawa} \& {Herbst}(1993{\natexlab{b}})}]{Hasegawa93a}
---. 1993{\natexlab{b}}, \mnras, 261, 83

\bibitem[{{Hincelin} {et~al.}(2015){Hincelin}, {Chang}, \&
  {Herbst}}]{Hincelin15}
{Hincelin}, U., {Chang}, Q., \& {Herbst}, E. 2015, \aap, 574, A24,
  \dodoi{10.1051/0004-6361/201424807}

\bibitem[{Hornek{\ae}r {et~al.}(2005)Hornek{\ae}r, Baurichter, Petrunin, Luntz,
  Kay, \& Al-Halabi}]{Hornekaer05}
Hornek{\ae}r, L., Baurichter, A., Petrunin, V.~V., {et~al.} 2005, The Journal
  of Chemical Physics, 122, 124701, \dodoi{10.1063/1.1874934}

\bibitem[{{Iqbal} \& {Wakelam}(2018)}]{Iqbal18}
{Iqbal}, W., \& {Wakelam}, V. 2018, \aap, 615, A20,
  \dodoi{10.1051/0004-6361/201732486}

\bibitem[{{Kalv{\={a}}ns} \& {Kalnin}(2019)}]{Kalvans19}
{Kalv{\={a}}ns}, J., \& {Kalnin}, J.~R. 2019, \mnras, 486, 2050,
  \dodoi{10.1093/mnras/stz1010}

\bibitem[{{Kalv{\={a}}ns} \& {Kalnin}(2020)}]{Kalvans20}
---. 2020, \aap, 641, A49, \dodoi{10.1051/0004-6361/202037906}

\bibitem[{{Kalv{\={a}}ns} \& {Kalnin}(2022)}]{Kalvans22b}
---. 2022, \apjs, 263, 5, \dodoi{10.3847/1538-4365/ac92e6}

\bibitem[{{Katz} {et~al.}(1999){Katz}, {Furman}, {Biham}, {Pirronello}, \&
  {Vidali}}]{Katz99}
{Katz}, N., {Furman}, I., {Biham}, O., {Pirronello}, V., \& {Vidali}, G. 1999,
  \apj, 522, 305, \dodoi{10.1086/307642}

\bibitem[{{Mart{\'\i}n-Dom{\'e}nech} {et~al.}(2024){Mart{\'\i}n-Dom{\'e}nech},
  {DelFranco}, {{\"O}berg}, \& {Rajappan}}]{Martin-Domenech24}
{Mart{\'\i}n-Dom{\'e}nech}, R., {DelFranco}, A., {{\"O}berg}, K.~I., \&
  {Rajappan}, M. 2024, \apj, 962, 107, \dodoi{10.3847/1538-4357/ad187e}

\bibitem[{{Mart{\'\i}n-Dom{\'e}nech} {et~al.}(2020){Mart{\'\i}n-Dom{\'e}nech},
  {Maksiutenko}, {{\"O}berg}, \& {Rajappan}}]{Martin-Domenech20}
{Mart{\'\i}n-Dom{\'e}nech}, R., {Maksiutenko}, P., {{\"O}berg}, K.~I., \&
  {Rajappan}, M. 2020, \apj, 902, 116, \dodoi{10.3847/1538-4357/abb59f}

\bibitem[{{Molpeceres} {et~al.}(2021){Molpeceres}, {Zaverkin}, {Watanabe}, \&
  {K{\"a}stner}}]{Molpeceres21}
{Molpeceres}, G., {Zaverkin}, V., {Watanabe}, N., \& {K{\"a}stner}, J. 2021,
  \aap, 648, A84, \dodoi{10.1051/0004-6361/202040023}

\bibitem[{{Neufeld} {et~al.}(2024){Neufeld}, {Welty}, {Ivlev}, {Caselli},
  {Edenhofer}, {Indriolo}, {Obolentseva}, {Silsbee}, {Sonnentrucker}, \&
  {Wolfire}}]{Neufeld24}
{Neufeld}, D.~A., {Welty}, D.~E., {Ivlev}, A.~V., {et~al.} 2024, \apj, 973,
  143, \dodoi{10.3847/1538-4357/ad7264}

\bibitem[{{Obolentseva} {et~al.}(2024){Obolentseva}, {Ivlev}, {Silsbee},
  {Neufeld}, {Caselli}, {Edenhofer}, {Indriolo}, {Bisbas}, \&
  {Lomeli}}]{Obolentseva24}
{Obolentseva}, M., {Ivlev}, A.~V., {Silsbee}, K., {et~al.} 2024, \apj, 973,
  142, \dodoi{10.3847/1538-4357/ad71ce}

\bibitem[{{Padovani} {et~al.}(2018){Padovani}, {Ivlev}, {Galli}, \&
  {Caselli}}]{Padovani18}
{Padovani}, M., {Ivlev}, A.~V., {Galli}, D., \& {Caselli}, P. 2018, \aap, 614,
  A111, \dodoi{10.1051/0004-6361/201732202}

\bibitem[{{Punanova} {et~al.}(2018){Punanova}, {Caselli}, {Pineda}, {Pon},
  {Tafalla}, {Hacar}, \& {Bizzocchi}}]{Punanova18}
{Punanova}, A., {Caselli}, P., {Pineda}, J.~E., {et~al.} 2018, \aap, 617, A27,
  \dodoi{10.1051/0004-6361/201731159}

\bibitem[{{Sandford} {et~al.}(1993){Sandford}, {Allamandola}, \&
  {Geballe}}]{Sandford93}
{Sandford}, S.~A., {Allamandola}, L.~J., \& {Geballe}, T.~R. 1993, Science,
  262, 400, \dodoi{10.1126/science.11542874}

\bibitem[{{Silsbee} {et~al.}(2021){Silsbee}, {Caselli}, \& {Ivlev}}]{Silsbee21}
{Silsbee}, K., {Caselli}, P., \& {Ivlev}, A.~V. 2021, \mnras, 507, 6205,
  \dodoi{10.1093/mnras/stab2546}

\bibitem[{{Sipil{\"a}} {et~al.}(2015){Sipil{\"a}}, {Caselli}, \&
  {Harju}}]{Sipila15a}
{Sipil{\"a}}, O., {Caselli}, P., \& {Harju}, J. 2015, \aap, 578, A55,
  \dodoi{10.1051/0004-6361/201424364}

\bibitem[{{Sipil{\"a}} {et~al.}(2019){Sipil{\"a}}, {Caselli}, {Redaelli},
  {Juvela}, \& {Bizzocchi}}]{Sipila19a}
{Sipil{\"a}}, O., {Caselli}, P., {Redaelli}, E., {Juvela}, M., \& {Bizzocchi},
  L. 2019, \mnras, 487, 1269, \dodoi{10.1093/mnras/stz1344}

\bibitem[{Sipil{\"a} {et~al.}(2021)Sipil{\"a}, Silsbee, \& Caselli}]{Sipila21}
Sipil{\"a}, O., Silsbee, K., \& Caselli, P. 2021, \apj, 922, 126,
  \dodoi{10.3847/1538-4357/ac23ce}

\bibitem[{{Sipil{\"a}} {et~al.}(2020){Sipil{\"a}}, {Zhao}, \&
  {Caselli}}]{Sipila20}
{Sipil{\"a}}, O., {Zhao}, B., \& {Caselli}, P. 2020, \aap, 640, A94,
  \dodoi{10.1051/0004-6361/202038353}

\bibitem[{{Taquet} {et~al.}(2014){Taquet}, {Charnley}, \&
  {Sipil{\"a}}}]{Taquet14}
{Taquet}, V., {Charnley}, S.~B., \& {Sipil{\"a}}, O. 2014, \apj, 791, 1,
  \dodoi{10.1088/0004-637X/791/1/1}

\bibitem[{{Vidali} {et~al.}(1991){Vidali}, {Ihm}, {Kim}, \& {Cole}}]{Vidali91}
{Vidali}, G., {Ihm}, G., {Kim}, H.-Y., \& {Cole}, M.~W. 1991, Surface Science
  Reports, 12, 135, \dodoi{10.1016/0167-5729(91)90012-M}

\bibitem[{{Wakelam} {et~al.}(2015){Wakelam}, {Loison}, {Herbst}, {Pavone},
  {Bergeat}, {B{\'e}roff}, {Chabot}, {Faure}, {Galli}, {Geppert}, {Gerlich},
  {Gratier}, {Harada}, {Hickson}, {Honvault}, {Klippenstein}, {Le Picard},
  {Nyman}, {Ruaud}, {Schlemmer}, {Sims}, {Talbi}, {Tennyson}, \&
  {Wester}}]{Wakelam15}
{Wakelam}, V., {Loison}, J.-C., {Herbst}, E., {et~al.} 2015, \apjs, 217, 20,
  \dodoi{10.1088/0067-0049/217/2/20}

\bibitem[{{Zhao} {et~al.}(2018){Zhao}, {Caselli}, \& {Li}}]{Zhao18}
{Zhao}, B., {Caselli}, P., \& {Li}, Z.-Y. 2018, \mnras, 478, 2723,
  \dodoi{10.1093/mnras/sty1165}

\end{thebibliography}

\appendix

\section{Calculation of the heating frequencies and deposited energy}\label{a:heatingFrequencyDerivation}

\begin{figure*}
\centering
\includegraphics[width = 0.7\textwidth]{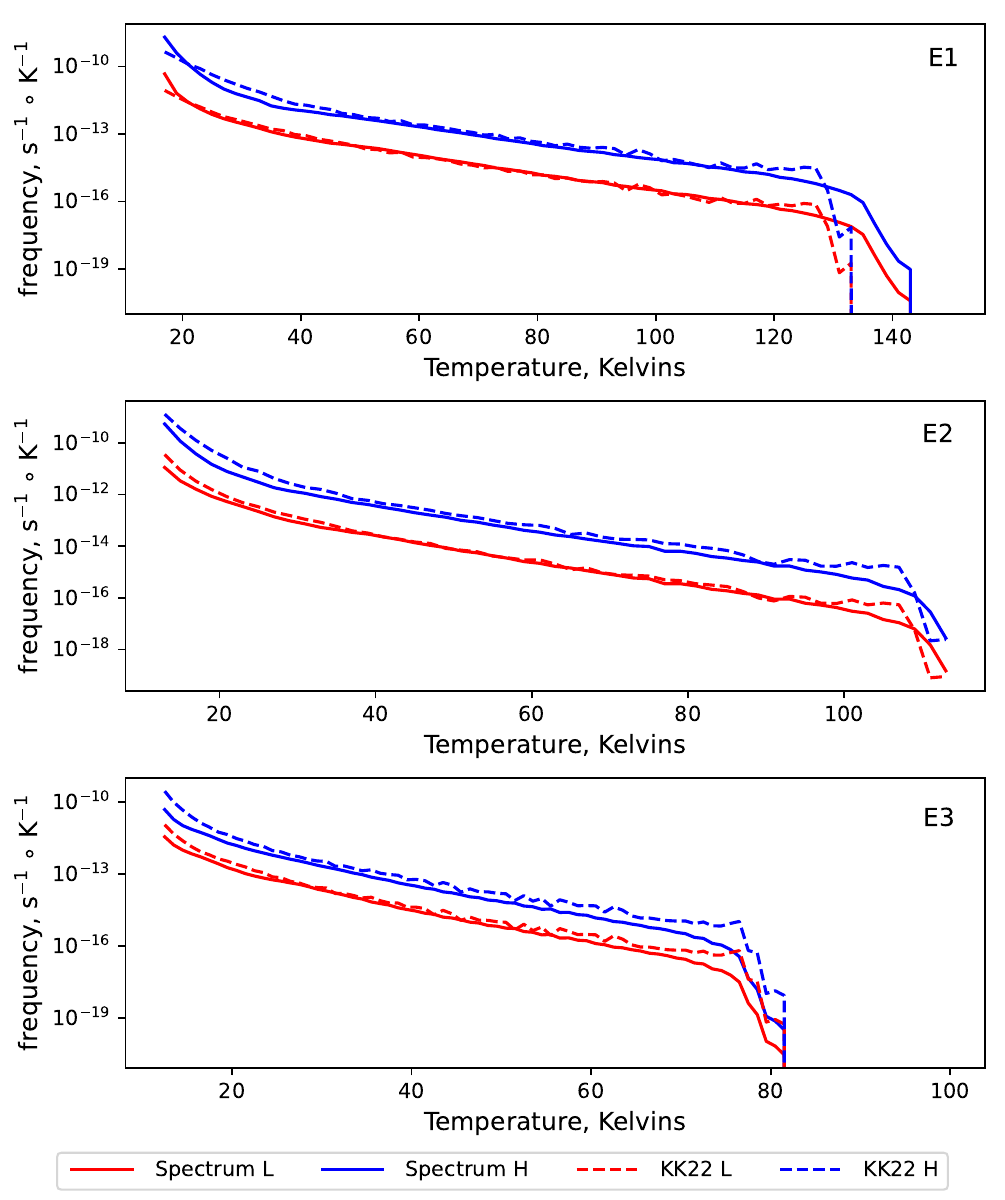}
\caption{Comparison of the heating frequencies given in \citet{Kalvans22b} with those we calculated.  The three panels correspond to the environments (mantle thickness and composition, and amount of shielding material) E1~to~E3 as described in the main text. In the legend, ``spectrum L'' and ``H'' refer to the P18 Low and High CR spectra, respectively, while KK22 refers to \citet{Kalvans22b}. We note that we assumed an initial grain temperature of 15 Kelvin for environment E1, as opposed to 10\,K assumed by \citet{Kalvans22b}, which is why our curves rise above those of \citet{Kalvans22b} for the lowest temperatures. }
 \label{fig:temperatureSpectra}
\end{figure*}

We verified the temperature spectra calculated by \citet{Kalvans22b} to the accuracy justified by the uncertainties in the CR spectrum.  Our calculations differ in several respects as listed below from those in \citet{Kalvans22b}:
\begin{enumerate}
\item{\citet{Kalvans22b} use the SRIM2013 package to determine the loss function for each cosmic ray species.  We used the loss function tabulated by Marco Padovani \citep{Padovani18} for protons.  For other species, we assumed the loss function (expressed as a function of energy per nucleon) to be that of protons multiplied by $Z^2/A$, where $Z$ and $A$ are the charge and mass number of the species respectively.}
\item{We were unable to locate heat capacities for CO above a temperature of 85 Kelvin in the reference cited in \citet{Kalvans22b}, and therefore we assume the heat capacity is constant above 85 Kelvin.  We note also that at atmospheric pressure, CO is observed to melt at 69 Kelvin, and boil at 81 Kelvin.  The latent heats associated with these transitions are not taken into account in either our model or that of \citet{Kalvans22b}.  
\item{\citet{Kalvans22b} used a variety of tracks through cuboid grains to account for the effect of impact parameter.  We considered spherical grains with the full range of CR impact parameters.}}
\end{enumerate}
A comparison between our calculations and those shown in \citet{Kalvans22b} is shown in Figure \ref{fig:temperatureSpectra}. Despite the somewhat different initial setup, we obtain frequency curves very close to those of \citet{Kalvans22b}, showing that the initial values used for the chemical simulations are robust.

\section{Results of simulations using the P18 High CR spectrum}\label{a:P18H}

\begin{figure*}
\centering
        \includegraphics[width=1.0\columnwidth]{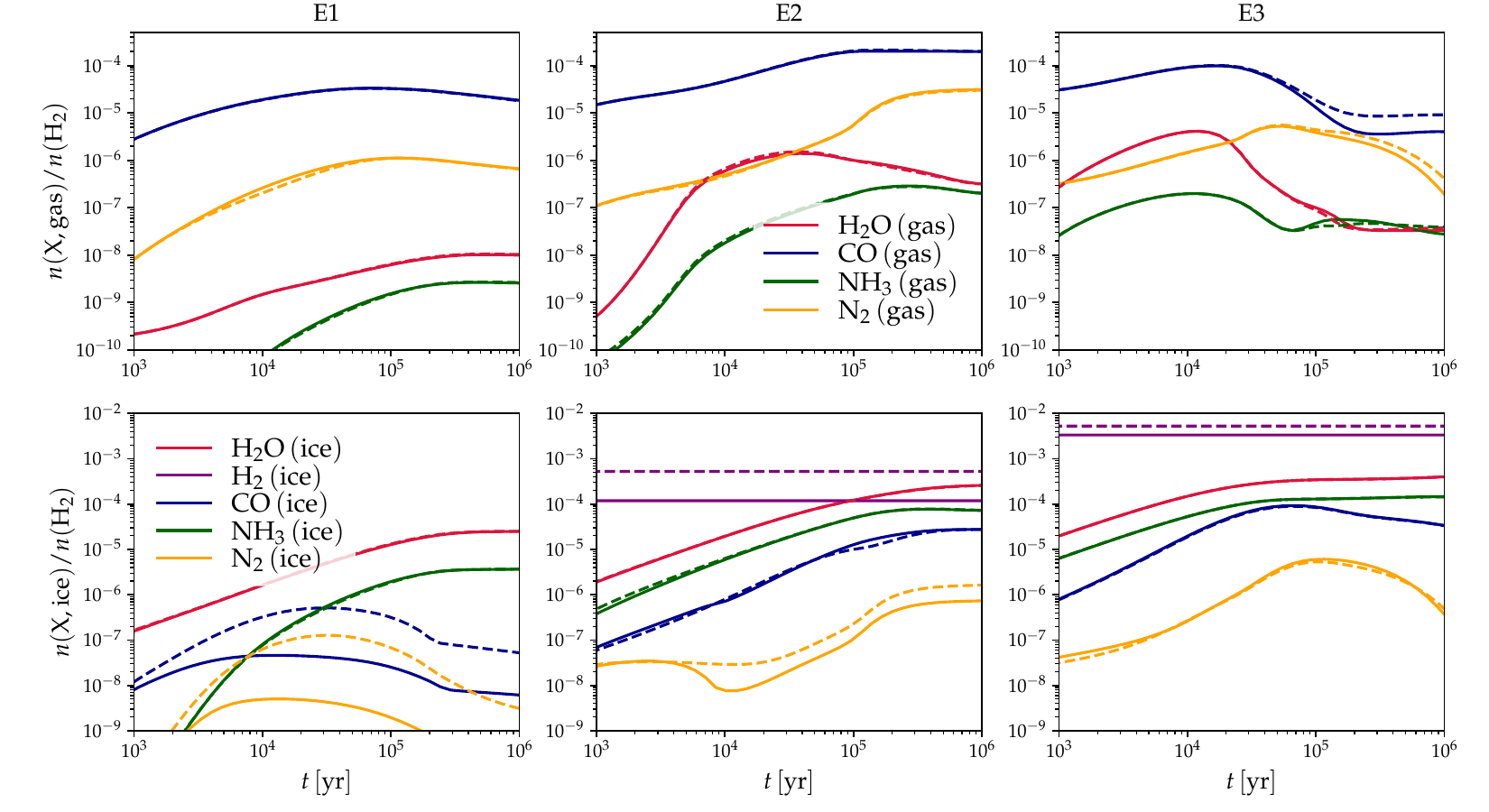}
    \caption{As Fig.\,\ref{fig:abundances}, but using the P18 High CR spectrum.}
    \label{fig:abundances_P18H}
\end{figure*}

As we have explained in the main text, the main reason for the inefficiency of the weak heating events in removing $\rm H_2$ from grain surfaces is that the CR strikes do not occur often enough. The frequency of the CR impacts is directly tied to the assumed CR flux. We show in Fig.\,\ref{fig:abundances_P18H} the analog of Fig.\,\ref{fig:abundances} in the main text, but calculated using heating frequencies based on the P18 High CR spectrum. We note that in this case the cosmic ray ionization rate is set to $\zeta = 10^{-16} \, \rm s^{-1}$ in the chemical model to take into account the increased ionization due to the higher CR flux, which also means that a direct comparison to Fig.\,\ref{fig:abundances} is not completely straightforward because the increase of the ionization rate from $1.3\times10^{-17} \, \rm s^{-1}$ to $10^{-16} \, \rm s^{-1}$ has an effect of its own on chemical evolution (increasing the ionization rate has an indirect effect on the grain cooling time). Regardless, Fig.\,\ref{fig:abundances_P18H} indicates that the higher CR flux does lead to a reduction in the abundance of $\rm H_2$ in the ice, but the effect is rather small and diminishes towards higher densities as expected based on the results presented in the main text. We conclude that weak heating events do not provide enough desorption to significantly affect the abundance of $\rm H_2$ in the ice -- at least as long as the grains are monodisperse and relatively large.

\end{document}